\documentclass[twocolumn,showpacs,superscriptaddress,preprintnumbers,amsmath,amssymb,aps,prl]{revtex4-1}
\usepackage{graphicx}
\usepackage{mathptmx}
\usepackage{verbatim}
\usepackage{graphicx}
\usepackage{amsmath}
\usepackage{amssymb}
\usepackage{amsfonts}
\usepackage{mathrsfs}
\usepackage{amsmath} 
\usepackage{color}
\usepackage{graphicx}
\usepackage{bm} 
\usepackage{amssymb}
\usepackage{xspace}
\usepackage{hyperref}
\usepackage{float}
\usepackage{bbold}
\usepackage{tabu}
\usepackage{textcomp}
\usepackage{dcolumn}
\newcommand{\sgn}{\text{sgn}}

\newcolumntype{L}[1]{>{\raggedright\arraybackslash}m{#1}}
\newcolumntype{C}[1]{>{\centering\arraybackslash}m{#1}}
\newcolumntype{R}[1]{>{\raggedleft\arraybackslash}m{#1}}
\newcolumntype{N}{@{}m{0pt}@{}}

\begin{document}

\title{Gate-Tunable Topological Flat Bands in Trilayer Graphene-Boron Nitride Moir\'e Superlattices}

\author{Bheema Lingam Chittari}
\affiliation{Department of Physics, University of Seoul, Seoul 02504, Korea}
\affiliation{Department of Physics, University of Carlifornia at Berkeley, Berkeley, CA 94709, USA}

\author{Guorui Chen}
\affiliation{Department of Physics, University of Carlifornia at Berkeley, Berkeley, CA 94709, USA}

\author{Yuanbo Zhang}
\affiliation{Institute for Nanoelectronic Devices and Quantum Computing, Fudan University, Shanghai 200433, China}

\author{Feng Wang}
\affiliation{Department of Physics, University of Carlifornia at Berkeley, Berkeley, CA 94709, USA}

\author{Jeil Jung}
\affiliation{Department of Physics, University of Seoul, Seoul 02504, Korea}
\affiliation{Department of Physics, University of Carlifornia at Berkeley, Berkeley, CA 94709, USA}

\begin{abstract}
We investigate the electronic structure of the flat bands induced by moir\'e superlattices 
and electric fields in nearly aligned ABC trilayer graphene-boron nitride interfaces 
where Coulomb effects can lead to correlated gapped phases. 
Our calculations indicate that valley-spin resolved isolated superlattice flat bands that 
carry a finite Chern number $C = 3$ proportional to layer number can appear near charge neutrality 
for appropriate perpendicular electric fields and twist angles.
When the degeneracy of the bands is lifted by Coulomb interactions these 
topological bands can lead to anomalous quantum Hall phases 
that embody orbital and spin magnetism. 
Narrow bandwidths of $\sim10$~meV achievable for a continuous range of twist angles $\theta \lesssim 0.6^{\circ}$ 
with moderate interlayer potential differences of $\sim$50~meV make the TLG/BN systems a promising platform for the 
study of electric-field tunable Coulomb interaction driven spontaneous Hall phases.
\end{abstract}
\maketitle

Generation of moir\'e superlattices in graphene and other 2D materials by forming 
van der Waals interfaces has emerged as an
efficient route to tailor high quality artificial band structures
~\cite{dean2013hofstadter,ponomarenko2013cloning,hunt2013massive}.
In particular, the periodic moir\'e patterns in the length scale of a few tens of nanometers 
that arise due to small lattice constant mismatch or twist angles
with the substrate give rise to moir\'e mini Brillouin zones 
whose zone corners are at energy ranges accessible by conventional field effects in gated transistor devices
~\cite{santos2007graphene,li2010observation,yankowitz2012emergence,bistritzer2011moire,wallbank2013generic,moon2012energy,jung2014ab}. 
The interlayer coupling become effectively strong in the limit of long moir\'e pattern periods
due to non-perturbative coupling between the superlattice zone folded moir\'e bands~\cite{bistritzer2011moire,jung2014ab},
which are suggestive that flat bands can routinely form in the limit of long moir\'e pattern periods
for a variety of 2D material combinations including 
twisted bilayer graphene and 
transition metal dichalcogenides heterojunctions~\cite{flatbandtmd2018allan,flatbandtmd2018randy,flatbandtmd2018kaxiras}.
Recent experiments have shown resistance peaks as a function of carrier doping indicative of Mott phases
in twisted bilayer graphene at the first magic twist angle~\cite{pjarillo_mott,pjarillo_superconducting} 
and in ABC trilayer graphene (TLG) nearly aligned with hexagonal boron nitride (BN)~\cite{guorui_mott2018}
when the Fermi energy is brought near the superlattice flat bands (SFB).  
In this work we carry out an analysis of the SFB in TLG/BN showing that they are generally topological bands, i. e. have finite Chern numbers, 
and the lifting of the valley-spin degeneracy by Coulomb interaction driven gaps of these Chern flat bands (CFB)
can give rise to quantum anomalous Hall phases with orbital and spin magnetism
even in the absence of an external magnetic field.
%

{\em Model Hamiltonian}$-$
The model Hamiltonian for  ABC stacked TLG is based on the low energy model for trilayer graphene with the band parameters
obtained from density functional theory local density approximation (LDA)~\cite{zhangfan_trilayer,koshino_trilayer,jjung_unpublished}.
We represent the Hamiltonian acting in the basis of the low energy sites A for bottom and B for top layers. 
Each band is four-fold degenerate with two-fold degeneracy in the principal valleys $(K, K')$ labeled with $\nu=\pm1$,
and two-fold degeneracy in real spin $(\uparrow, \downarrow)$ labeled with $s=\pm1$.
We label the lowest valence and conduction $(h,e)$ bands through $b =\pm1$.  
The low energy Hamiltonian for a rhombohedral $N$-layer graphene is 
\begin{equation}
\label{eq1}
{H}_{N}^{\nu, \xi}
= 
\frac{\upsilon_0^N}{(-t_1)^{ N-1}}\left(
\begin{array}{cc}
0 & \left( {\pi }^{\dag }\right) ^{N}     \\
{\pi ^{N}} & 0
\end{array} \right) 
+ \Delta \sigma_z
+  H^{\rm R}_{N} + H^{\rm M}_{\xi}, 
\end{equation}
where $\pi =  (\nu p_x + i p_y)$. 
We will discuss our results for $\nu=1$  principal valley $K$ unless stated otherwise. 
The parameter $\Delta$ represents an adjustable interlayer potential difference between the top and 
bottom layers that include the effects of a perpendicular electric field and its screening.
In a TLG with $N=3$ layers we model the remote hopping term corrections through
%
%
\begin{eqnarray}
\label{eq2}
H^{\rm  R}_{3, \nu} &=&
\left[ 
\left(
\frac{2 \upsilon_0 \upsilon_3 p^2}{t_1} + t_2
\right)   \sigma_{x} 
 \right]
\\
&+&
 \left[  \frac{2 \upsilon_0 \upsilon_4 p^2}{t_1^{ 2}}  - \Delta'  + \Delta'' \left(
1- \frac{3\upsilon_0^2p^2}{t_1^{ 2}}
\right)  \right]  \mathbb{1}.
\nonumber  
\end{eqnarray}
The effective hopping parameters are $t_0=-2.62$~eV, 
$t_1= 0.358$~eV, $t_2=-0.0083$~eV, $t_3=0.293$~eV and $t_4=-0.144$~eV ~\cite{accuratebilayer,jjung_unpublished}, 
where the associated velocities are defined as $\upsilon_m = \sqrt{3} a \, | t_m |  / 2\hbar$
with $a=2.46\,\AA$ the lattice constant of graphene.
The diagonal terms $\Delta' = 0.0122$~eV, and $\Delta'' = 0.0095$~eV are used to provide an 
accurate fit for the LDA bands. 
The moir\'e patterns have a period $\ell_M \simeq a / (\varepsilon^2 + \theta)^{1/2}$
that depends on $\varepsilon = (a - a_{BN})/a_{BN}$ the relative lattice constant mismatch between graphene and BN, and  the twist angle $\theta$.
The moir\'e potential generated in graphene due to BN is given by
\begin{equation}
H^M_{\xi=\pm1} (\vec{r}) = V_{A/B}^M (\vec{r}) = 2 C_{A/B} {\rm Re} \left[ e^{i \phi_{A/B}} f(\vec{r}) \right] 
\frac{( \xi \sigma_z + \mathbb{1} )}{2}
\end{equation} 
where the moir\'e parameters are
$C_A = -14.88$~meV, $\phi_A = 50.19^{\circ}$ and $C_B = 12.09$~meV, 
$\phi_B = -46.64^{\circ}$~\cite{jung2014ab,jung2015origin}, and  the auxiliary function
$f(\vec{r}) = \sum_{m=1}^6 e^{i \vec{G}_m \cdot\vec{r}} (1 + (-1)^m )/2$ is expressed using the six moir\'e reciprocal lattice 
$\vec{G}_{m=1 \dots 6} = \hat{R}_{2\pi (m-1)/3} \vec{G}_1$ successively rotated by sixty degrees where $\vec{G}_1=(0, 4\pi / (\sqrt{3} \ell_M ))$.
The parameter $\xi = \pm 1$ distinguishes the two types of possible alignments between TLG and BN
where we have defined the moir\'e pattern to perturb only the low energy A (bottom) or B (top) sites in the graphene layer contacting BN.
The two moire patterns will give rise to rather different flat band behaviors as a function of $\Delta$ and twist angle.
These moir\'e pattern parameters given in sublattice basis can be related with pseudospin basis parameters $C_0 = -10.13$~meV, $\phi_0 = 86.53^{\circ}$, 
and $C_z = -9.01$~meV, $\phi_z = 8.43^{\circ}$~\cite{jung2014ab,laksono2017}.

\begin{figure}
\includegraphics[width=8cm]{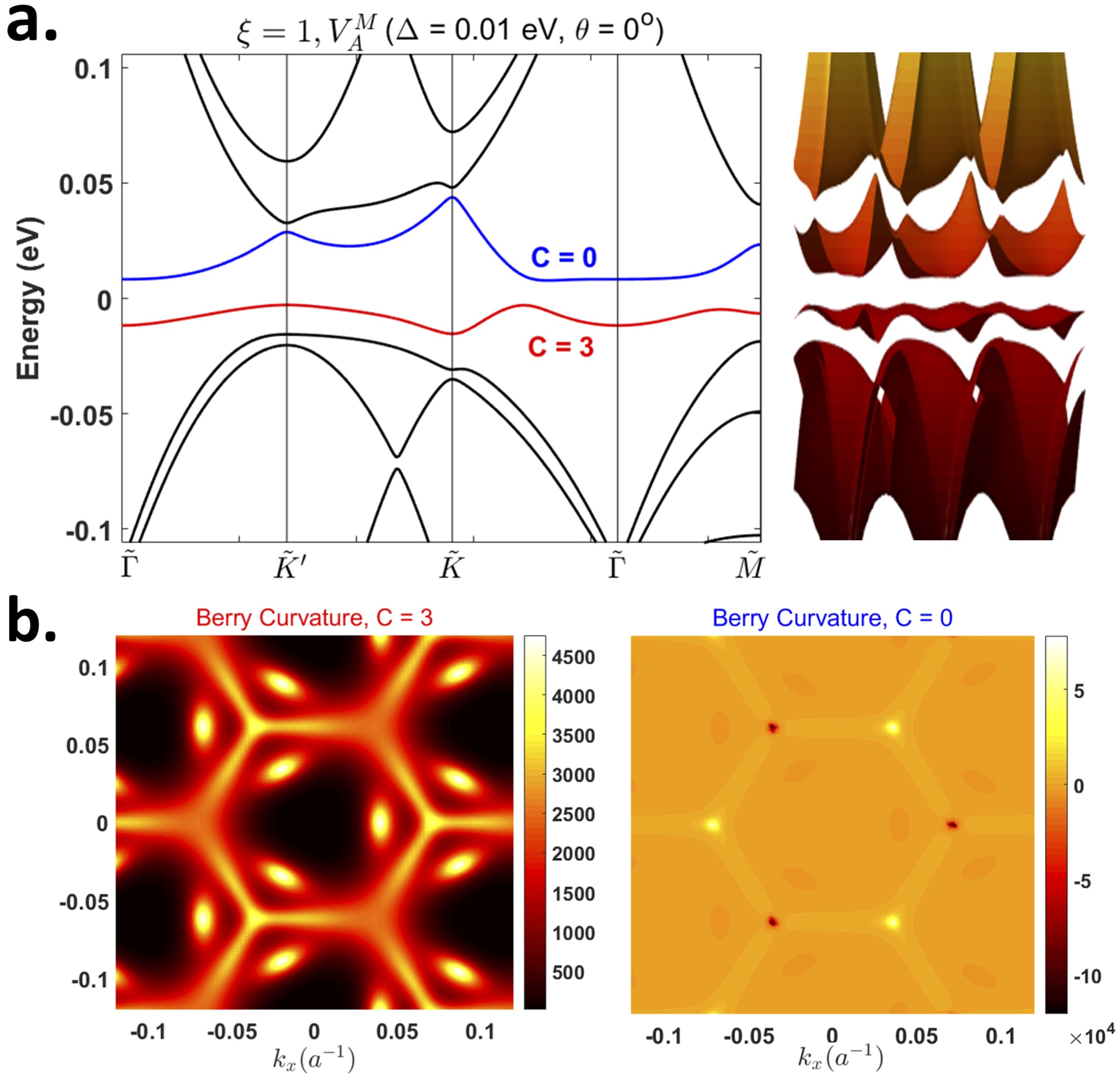} 
\vspace{-15pt}
\begin{center}
\end{center}
 \caption{
 (Color online)  
 {\bf a.} The band structure of ABC-TLG for zero twist angle near charge neutrality in the folded zones representation 
 subject to $V^{M}_A=H^{M}_{\xi=1}$  
 moir\'e patterns and interlayer 
 potential differences of $\Delta = 10$~meV  that give rise to flat Chern bands with $C= 3$ represented in red and $C=0$ trivial
 band represented in blue.
 {\bf b.} The Berry curvatures for the valence and conduction band structures of panel (a) where we see  
Berry curvature hotspots near the trigonal warping band edges and mBZ zone boundaries that add up 
 in the Chern band. In the trivial band we see sharp peaks with opposite Berry curvatures mainly at the mBZ that cancel out.
}
 \vspace{-12pt} 
 \label{figure2}
\end{figure}

\begin{figure*}
\begin{center}
\includegraphics[width=17cm]{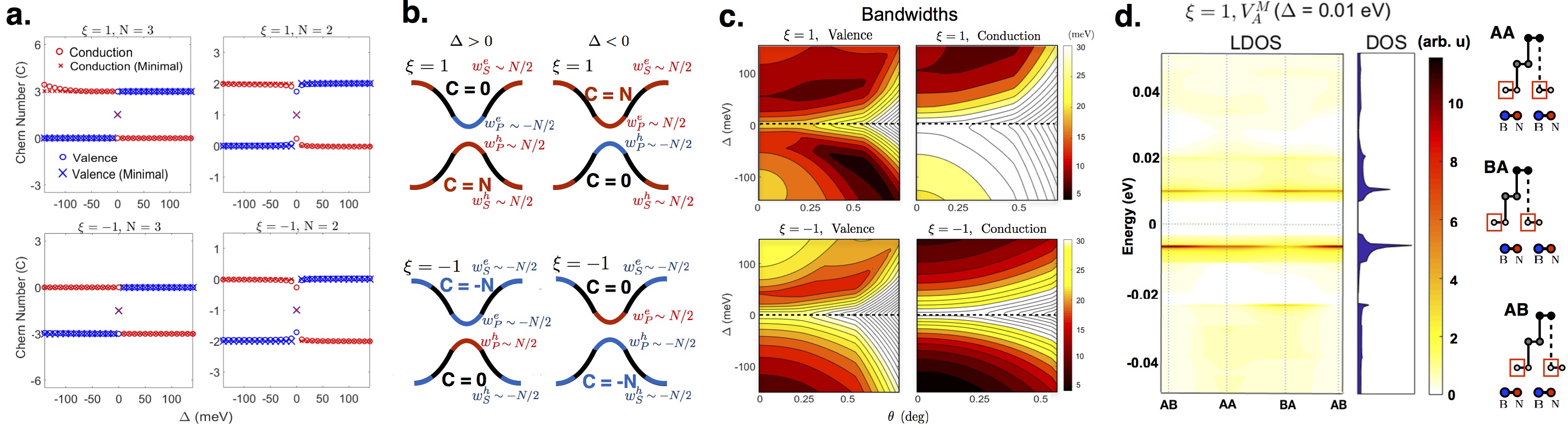}
\vspace{-15pt}
\end{center}
 \caption{
 (Color online) 
 {\bf a.} Chern numbers for the valence and conduction flat bands for aligned TLG on BN for the bands in Eqs.~\ref{eq1}-\ref{eq2},
 bilayer on BN using $H^R_2$ from Ref.~\onlinecite{accuratebilayer} and corresponding minimal $H^{\rm R}_{\xi} = 0$ models, 
 calculated using a total of 18361 $k$-points in the mBZ. An abrupt transition happens when the sign of $\Delta$ changes near zero field. 
 {\bf b.} Schematic illustration 
 of the Chern weights near $\tilde{\Gamma}$ and mBZ boundaries in the limit of $\left| \Delta \right| \ll 1$ that add up into an integer $C^{e/h}= w^{e/h}_P + w^{e/h}_S$, where $w^e_P= - w^h_P$ and $w^e_S=w^h_S$ so that either the electron or hole band has  a finite Chern number. 
 {\bf c.} Bandwidth phase diagram for the valence and conduction bands as a function of interlayer potential difference $\Delta$
 and twist angle $\theta$ calculated from the difference between the maximum and minimum eigenvalue within a band. 
 {\bf d.} Local stacking and energy dependent  LDOS and DOS manifesting the localization of flatband wavefunctions. 
  }
 \vspace{-12pt} 
 \label{figure3}
\end{figure*}

\smallskip

{\em Topological flat bands in TLG/BN superlattices}$-$
The presence of moir\'e superlattices can produce avoided gaps at the moir\'e mini Brilloun 
zone (mBZ) boundaries~\cite{dasilva2015,shuyun,laksono2017}
while an additional gap at the primary Dirac point would isolate the low energy 
bands near charge neutrality~\cite{song_pnas, jung2015origin,guorui_mott2018}.
Because the low energy bands in ABC trilayers are less dispersive than those of single or 
bilayer graphene, they are particularly suitable for the efficient isolation and 
narrowing of the low energy bands by electric fields that enhance both the primary 
and secondary gaps~\cite{guorui_mott2018}. 
Here, we discuss how the gaps opened by perpendicular electric fields lead to Berry curvatures in 
the isolated flat bands of TLG/BN and can turn them into topological Chern bands with a quantized Hall effect. 
The Berry curvature for the $n^{th}$ band can be
calculated using the standard formula 
$\Omega_n (\vec{k}) = -2 \sum_{n' \neq n} Im \left[  \langle u_n | \frac{\partial H}{ \partial k_x} | u_{n'} \rangle    \langle u_{n'} |  \frac{\partial H}{ \partial k_y} | u_{n} \rangle 
/ \left( E_{n'} - E_n \right)^2  \right]$
%
%
~\cite{rmp_berry}
where for every $k$-point we take sums through all the neighboring $n'$ bands, 
the $| u_n \rangle$ are the moir\'e superlattice Bloch states and $E_n$ are the eigenvalues.
The Chern number of the $n^{th}$ band is obtained from the Berry curvature through
$C = \int {\rm d^2} \vec{k} \,\, \Omega_{n}(\vec{k})/(2\pi)$ integrated in the Brillouin zone.
Here we use an abbreviated notation $C = C_{\nu,s,N,\xi, b}$ with implicit valley, spin indices
and other system parameters.
In the band structures of TLG/BN in Fig.~\ref{figure2} we can clearly observe the primary gap near charge 
neutrality and secondary gaps near the mBZ boundaries as well as the Berry curvature hotspots that 
illustrate the distribution of Chern number weights, often near the mBZ boundaries where the secondary gaps open. 

Our numerical calculations for TLG/BN predict the Chern numbers $C = \pm3$ for the CFBs depending on system parameters.
We have verified up to trilayers that 
\begin{equation}
C = N \nu \xi  \delta_{\sgn(\Delta) \cdot \xi,b}
\end{equation}
in an $N$-chiral graphene two dimensional electron gas (2DEG),
where the Chern bands are found for either the valence or conduction bands depending on applied electric 
field sign and moir\'e pattern potential.
In Fig.~\ref{figure3} we show that the calculated Chern numbers are quantized for a wide range of $\Delta$ values
for valence and conduction bands of TLG/BN and BLG/BN.
In particular we find that either the conduction or valence SFB near charge neutrality 
becomes a Chern band as soon as $\Delta$ opens a gap at the primary Dirac point.
This behavior can be understood if we consider that a rhombohedral 
$N$-layer graphene~\cite{hongki2008} develops in the limit of small $\Delta$
a primary Chern weight $w_{P} \sim \sgn(\Delta ) b \nu \, N/2$ near each valley
 whose sign depends on $\sgn(\Delta)$, the hole or electron band character 
$b=\pm1$, 
as well as valley $\nu=\pm 1$~\cite{xiaodi2007,valleykink2011,zhangfan2011,nanoroads}.
In the absence of secondary gaps due to moire patterns the $K$ and $K^{\prime}$ principal valleys are mutually connected and 
the valley Chern numbers identified as Chern weights $w_P$ near each valley are not protected topological numbers. 
Nevertheless, they give an intuitive idea about the Hall conductivity dynamics near the chiral 2DEG band edges 
and are useful for counting the number of zero-line modes in the valley Hall domain 
walls~\cite{volovik,martin,valleykink2011,zhangfan_pnas,changheelee,nanoroads,zhujun,julong}.
The situation is quite different when primary and secondary gaps are simultaneously present 
near the mBZ boundaries because the Chern weights sum $C^{e/h} = w^{e/h}_P + w^{e/h}_S$
needs to add up to a zero or finite integer value in each isolated band~\cite{hasan_kane_rmp}, where $w^{e}_S$ or $w^{h}_S$ are the 
secondary Chern weights for electron and hole bands.
The secondary Chern weights near the mBZ boundaries  
depend on the moir\'e pattern potentials that generate the avoided secondary gaps,
as evidenced by the fact that different moir\'e potentials $V^M_A$ or $V^{M}_B$ give rise to
flat bands with different Chern numbers. 
The abrupt change in the band Chern number with the sign of $\Delta$ can be related with the sign changes of $w_P$. 
Considering that $w^h_P = - w^e_P$
for electrons and holes in the limit $\left| \Delta \right| \ll 1$ we conclude that the 
secondary weights should initially be equal $w^{e}_S = w^{h}_S$, see Fig.~\ref{figure3},
while unequal electron-hole secondary Chern weights $w^e_S \neq w^h_S$ 
might be achievable using different moir\'e and Hamiltonian parameters. 
%
%
%
%
%
For increasing $\Delta$ the primary Chern weights are progressively pushed from 
the vicinity of $\tilde{\Gamma}$ towards the mBZ boundaries closer to the location of the secondary weights $w^{e/h}_{S}$ 
while maintaining a constant Chern number in the isolated bands. 
Conclusions similar to our analysis in TLG/BN could be expected in other rhombohedral-multilayer graphene 
$N$-chiral 2DEGs, although the remote hopping terms in the band Hamitonian are important to properly account 
for the flatband dispersions and the secondary gaps. 
The Supplemental material Figs.~S1-S2 for the Berry curvatures in minimal $N$-chiral 
multilayer graphene with $N$=1,2,3 illustrates the evolution of Chern weight distribution in the mBZ from small to large $\Delta$.

%

%
%

{\em Field-dependent bandwiths and localization-}
The external electric field strength modulates the size of the primary band gap near $\tilde{\Gamma}$ 
which impacts directly the shape of the low energy 
SFB and their optimum flatness will depend on field direction and strength. 
The non-monotonic behavior of the band flatness as a function of electric field strength 
highlights the electron-hole asymmetry inherent in TLG and the relevance 
of the moir\'e potential details in determining the band shapes. 
In Fig.~\ref{figure3} we represent a colormap that summarizes the evolution of the bandwidth 
of the SFB as a function of electric field and twist angle quantified through the difference 
between the maximum and minimum energy values within a given band.
We can observe that for every given twist angle there is often an optimum interlayer potential difference that maximizes
the band flatness either for positive and negative field directions, and that the overall flatness does not always grow 
monotonically with increasing electric field magnitude. 
Increasing $\Delta$ to appropriately large values will favor the onset of Coulomb interaction driven gaps
by increasing the separation of the SFB with neighboring energy bands and reducing its bandwidth.
%
In addition to the electric field magnitude and direction, the relative twist angle in the system has an impact in the flatness and energy location of the SFB. 
Introducing a finite twist angle is expected to widen the moir\'e bands due to the 
increase in the size of the mBZ in reciprocal space for reduced moir\'e real space periods.
However, for relatively small twist angle values of up to $\sim1^{\circ}$ the moir\'e pattern periods are still $\sim$10~nm 
and the enhanced suppression of achievable bandwidth through external electric fields can sufficiently compensate for this bandwidth increase due to rotation.
Direct information on the band flatness is reflected in the density of states (DOS) plots as a function of energy
allowing to find the energy regions where Coulomb correlations are expected to be stronger.
The local density of states (LDOS) plots for every local stacking configuration also provides
insights on the electron localization properties of the SFB electrons in real space, see Fig.~\ref{figure3}.
Charge and spin density modulations are expected to take place mainly around the LDOS peaks 
when the degeneracy of the bands is lifted due to Coulomb interactions. The LDOS plots provide valuable
information for guiding scanning probe microscopy experiments that search for flat band signatures.

\begin{figure}
\begin{center}
\includegraphics[width=7.5cm]{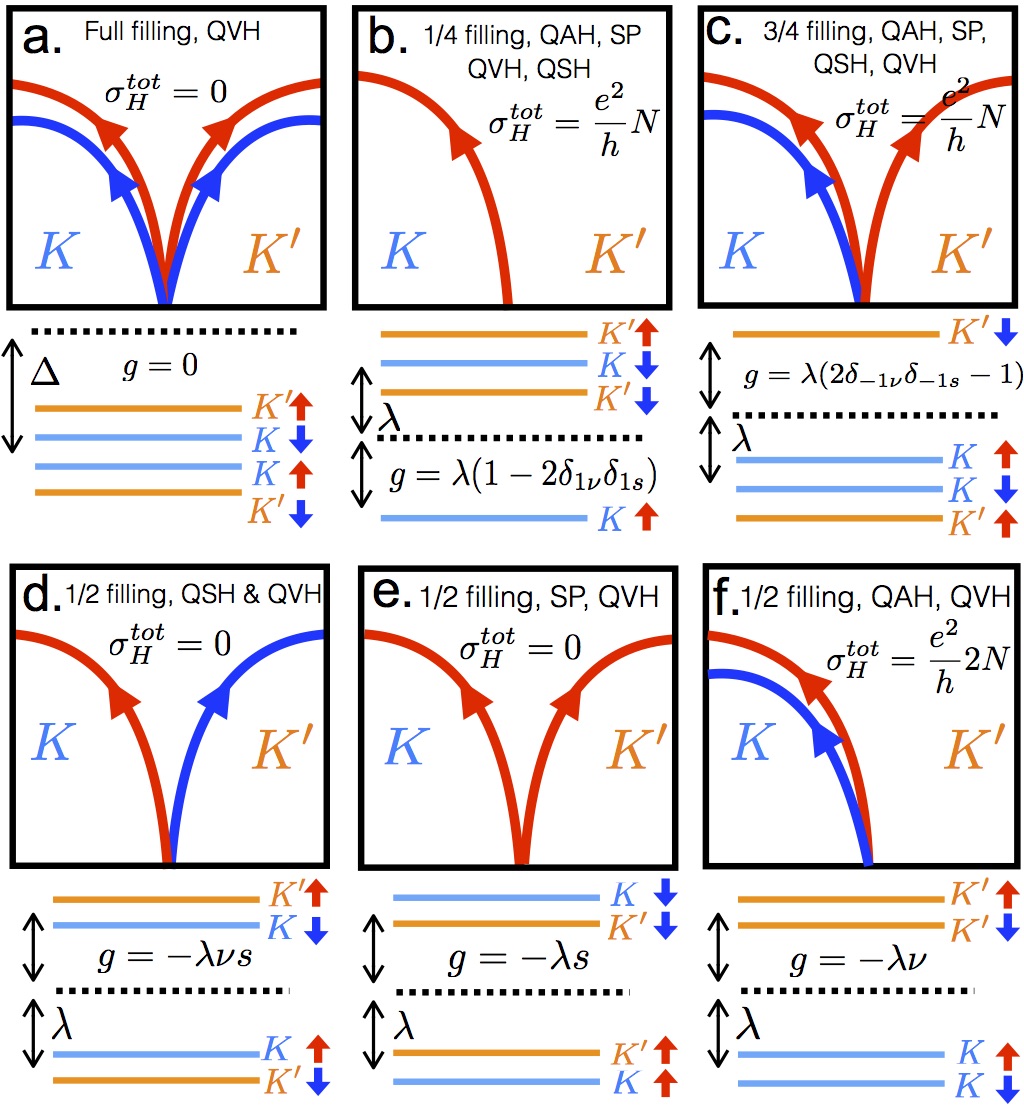}
\vspace{-15pt}
\end{center}
 \caption{
 (Color online)  
For the valence flat bands associated with $\Delta > 0$ and $\xi=1$ moir\'e potentials we schematically represent 
how the occupation of valley $(K, K^\prime)$ and spin $(\uparrow, \downarrow)$
resolved CFB contributes towards the generation orbital and spin magnetism.
Four different configurations of valley spin are possible for 1/4, 3/4 filling (b.-c.), and two per each 1/2 filling represented (d.-f.). 
The total charge Hall conductivity $\sigma^{tot}_H = \sum_i \sigma^i_H$ of the bands is contributed by $\sigma^{i}_{H} = C_{i} e^2/h$
for each occupied valley-spin flavor $i$  where $C_{i} = N \nu \xi \delta_{\sgn(\Delta) \cdot \xi, b}$, and the valley Hall conductivity 
is proportional to filling.
A variety of spontaneous quantum anomalous, valley and spin Hall effects 
should be expected when interaction driven gaps open for 1/4, 1/2, 3/4 filling of the CFB.
In particular 1/4 and 3/4 fillings are found to simultaneously have spin and orbital magnetism. 
 }
 \label{figure5}
\vspace{-12pt} 
\label{BS_Caa}
\end{figure}

{\em Spontaneous quantum Hall phases$-$} 
Filling of nontrivial Chern bands will lead to an associated quantum Hall effect
that should be observable in transport experiments.
An interesting scenario is found in partially filled CFB
when valley and spin degeneracy is lifted by Coulomb interactions 
giving rise to a variety of spontaneous quantum Hall phases.
We take as a working assumption that the interaction driven gapped states at 1/4, 1/2, 3/4 partial 
filling of the flat bands will develop into spin-collinear and valley-collinear phases where each
valley-spin flavor can be filled sequentially. 
In the Hartree-Fock approximation it is reasonable to assume
that same spin polarization will be preferred over non-collinear spin states as in conventional
quantum Hall ferromagnetism of Landau levels~\cite{qhf}, 
while valley polarized phases can be preferred due to momentum space exchange condensation 
over valley-coherent phases with two partially filled valleys~\cite{valleypolarized}. 
We represent Coulomb interaction driven gaps through rigid shifts proportional to $\lambda$ 
in the CFB energies through a function $g(\nu,s)$ to classify the different possible states
assuming any of them are possible.
We will follow a classification scheme closely similar to the four valley-spin components 
of $N$-chiral multilayer graphene in Refs.~\cite{zhangfan2011,jung2011}, 
in our case facilitated by the fact that the quantum valley Hall effect is proportional to CFB filling.
In Fig.~\ref{figure5} we illustrate the example of $b=1$, $\xi=1$, $\Delta >0$ case corresponding
to the top panel of Fig.~\ref{figure2} a representative selection of quantum Hall ground states
also summarized in Table~\ref{hallcond}.
%
\begin{table}[t]
\caption{ Summary of the Chern flat band configurations (1 for occupied, 0 for unoccupied) 
and corresponding charge, spin, and valley Hall
conductivities (in $e^2/h$ units) and insulator types: quantum anomalous Hall (QAH), 
spin Hall (QSH), valley Hall (QVH), and spin polarized (SP).
The layer number $N=3$ is equal to the flatband Chern number magnitude in TLG/BN.  }
\newcommand\T{\rule{0pt}{3.1ex}}
\newcommand\B{\rule[-1.7ex]{0pt}{0pt}}
\begin{scriptsize}
\centering
\begin{tabular}{c | c c c c || c | c | c |  c}
\hline\hline Fig. & $K\uparrow$ & $K\downarrow$ & $K'\uparrow$ & $K'\downarrow$ &
$\sigma^{\rm (CH)}$ & $\sigma^{\rm (SH)}$ & $\sigma^{\rm (VH)}$ &   Insulator\T\\[3pt]
\hline
\ref{figure5}(a)  & 1 & 1 & 1 & 1  & $0$   &$0$  &  $4N$ &   QVH \T \\[3pt]
\ref{figure5}(b)  & 1 & 0 & 0 & 0  &  $N$  &$0$  &  $N$ &   QAH, SP, QVH \T \\[3pt]
\ref{figure5}(c)  & 1 & 1 & 1 & 0  &  $N$  &$-N$ & $3N$     &   QAH, QSH, SP, QVH \T \\[3pt]
\ref{figure5}(d)  & 1 & 0 & 0 &  1  &  $0$  &$2N$ & $2N$ &   QSH, QVH \T \\[3pt]
\ref{figure5}(e)  & 1 & 0 & 1 &  0  &  $0$  &$0$ & $2N$ &   SP, QVH \T \\[3pt]
\ref{figure5}(f)   & 1 & 1 & 0 &  0  &  $2N$  &$0$ & $2N$ &  QAH, QVH\T \\[3pt]
\hline\hline
\end{tabular}
\end{scriptsize}
\label{hallcond}
\end{table}
At charge neutrality when all CFB are filled at filling 1 we have a valley Hall state where
the charge Hall conductivity summed over all occupied flavors 
totals to zero the charge Hall conductivity is $\sigma_H^{tot}= \sum_i \sigma^i_{H} = 0 $. 
The cases of 1/4 and 3/4 filling can be pictured through selective filling/emptying of 
a given $(\nu',s')$ band using shifts of 
$g(\nu,s) = \lambda \left(1 - 2 \delta_{\nu' \nu} \delta_{s' s} \right)$ and $\lambda \left(2 \delta_{\nu' \nu} \delta_{s' s} - 1\right)$ respectively. 
These are interesting cases with $\sigma_H^{tot} = \pm N e^2/h$ charge Hall conductivity
where we have simultaneously a quantum anomalous and spin Hall effect.
For the 1/2 filling when two CFB are filled we have a greater variety of quantum Hall ground states. 
One of the possibilities is the quantum anomalous Hall phase where two equal $\nu'$ valley CFBs are filled. 
The shift functions can then be modeled through $g (\nu,s) = \lambda (1-2\delta_{\nu' \nu})$ 
and the charge Hall conductivity is $\sigma_H^{tot}= 2 N \nu' e^2/h$.
The remaining two scenarios have Hall conductivity $\sigma_H^{tot}= 0 $, 
in the case where the same spin $s'$ are polarized
and are modeled through $g (\nu,s) = \lambda (1- 2 \delta_{s' s})$ shifts, and the 
quantum spin phase can be modeled through $g (\nu,s) = \pm \lambda \nu s$ shifts
depending on the relative signs of the occupied valley-spin indices.

{\em Discussions$-$} In this work we have analyzed the topological character of the 
superlattice flat bands (SFB) in ABC trilayer graphene-hexagonal boron nitride superlattices where 
signatures of gate tunable Mott gaps have been observed recently in experiments~\cite{guorui_mott2018}. 
Our analysis indicates that topological flat bands with Chern number $C = \pm 3$
will form either for electrons or holes depending on the electric field sign and moir\'e potential.
This scenario makes it possible to study Coulomb interaction driven ordered phases 
in zero and finite Chern number flatbands within the same device by modifying the carrier density
from electrons to holes, making this system an interesting platform for exploring 
the interplay of correlation physics with topological order.
Band gap openings for partial filling of the flat bands indicate that
a selective occupation of Chern flat bands (CFB) of different valley-spin flavors
should be possible. 
Assuming valley-spin collinear ground states of these partially filled CFB, 
different types of spontaneous quantum Hall phases with orbital and spin
magnetization can be expected, with total charge
Hall conductivities of zero or $\sigma^{tot}_H = \pm 6 e^2/h$ expected for 1/2 filling, 
whereas $\sigma^{tot}_H = \pm 3 e^2/h$ that is always finite
is expected for 1/4 or 3/4 filling. 
From a device application point of view, one important advantage of the
field tunable gapped Dirac materials
is that the bandwidth variations of the SFB are less sensitive to twist angle 
compared to twisted bilayer graphene where a precise twist angle control is required.

\begin{acknowledgments}
{\em Acknowledgments$-$}
We thank A. H. MacDonald for helpful discussions.
The initial idea and proof of principle calculation of 2D flatband engineering was supported by an ARO MURI award (W911NF- 15-1-0447). Y.Z. acknowledge financial support from National Key Research Program of China (2016YFA0300703), and NSF of China (grant nos. U1732274, 11527805 and 11425415).  Support from the Korean NRF is acknowledged for  B.L.C. through NRF-2017R1D1A1B03035932 and J.J. through NRF-2016R1A2B4010105.
\end{acknowledgments}
{\em Note added:} During the finalization of this manuscript we became aware of a related recent work \cite{senthil}.

\maketitle
\begin{center}
{\bf \large Supplemental Material}
\end{center}

In this supplement we present band sructures and Berry curvatures that are obtained for the minimal band Hamiltonian for trilayer, bilayer and monolayer graphene on boron-nitride modeled by Eq.~1 in the main text setting the remote hopping terms to zero $H^R_N=0$ and subject to sublattice selective moire potentials 
$H^M_{\xi}$ labeled with $\xi=\pm1$.
We present results for the Berry curvatures in the limit of small $\Delta$ = $\pm$  0.0001 and  moderately large $\Delta = 0.1$~eV are clearly shown in the Fig.~S1 and S2 respectively to illustrate the distribution of the Chern weights depending on primary gap size. The Fig.~S1 shows the Berry curvatures alone for small gaps, wheres the Fig.~S2 shows the band structures together with the berry curvatures for the larger gaps where we can visualize the redistribution of the Chern weights.
Fig.~S3 illustrates the band structure and Berry curvatures of graphene on boron nitride subject 
to realistic moir\'e patterns from (PRB 89, 205414 (2014)) that include both the diagonal terms and off-diagonal components, 
plus an {\em ad hoc} gap at the primary Dirac point to verify the formation of isolated bands.
\begin{figure*}
\begin{center}
\includegraphics[width=18cm]{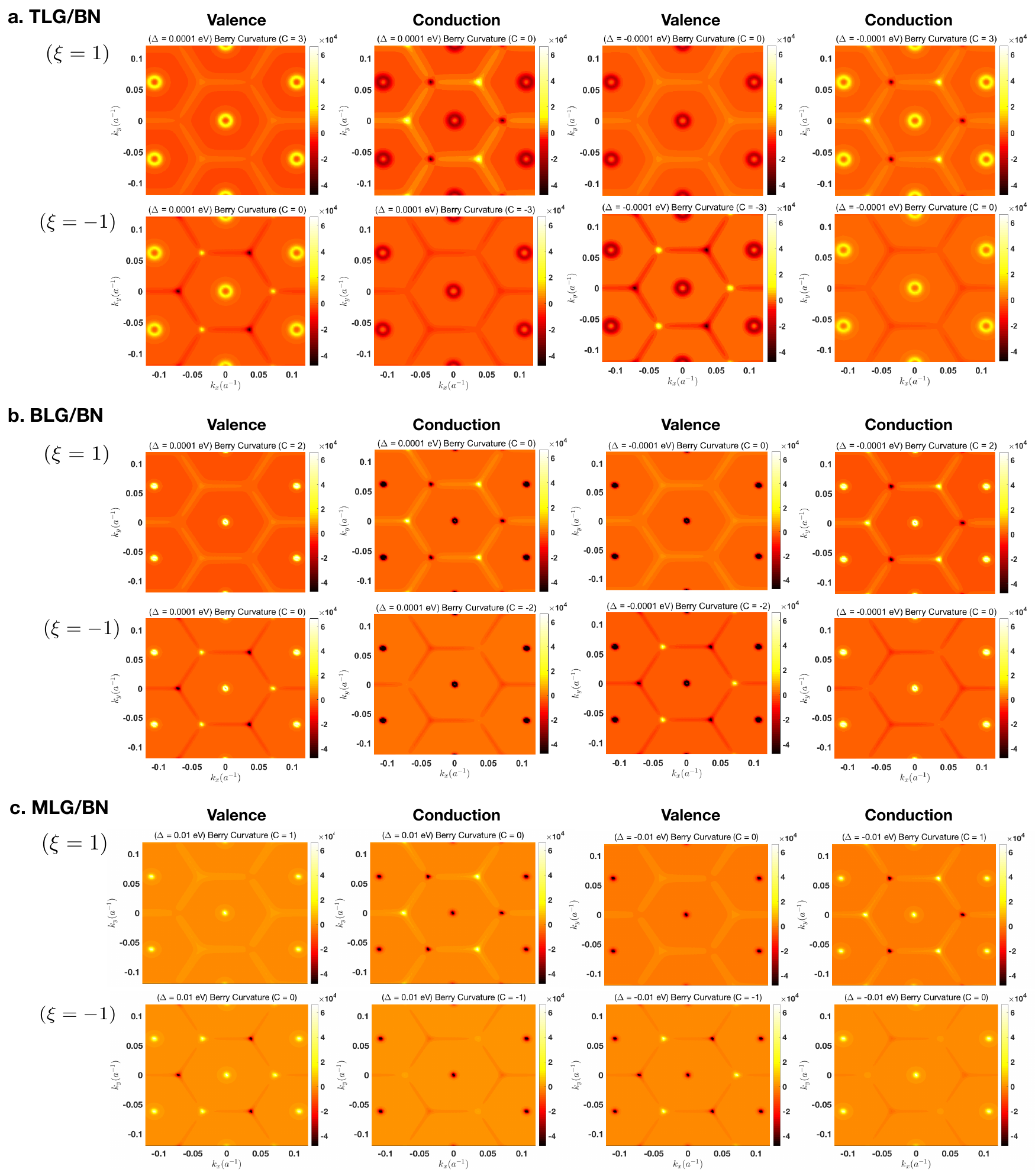}
\end{center}
 \caption{
 (Color online) The Berry curvatures of ABC-TLG ({\bf a. TLG/BN}), bilayer ({\bf b. BLG/BN}) and monolayer ({\bf c. MLG/BN}) 
 peaked near $\tilde{\Gamma}$ in the limit of small interlayer potential 
 differences $\Delta$ associated to the valence band (left) and conduction bands (right) near
  charge neutrality  of rhombohedral $N$-chiral graphene on BN modeled by Eq.~1 in the main 
  text for N=3,2,1 with $H^{R}_N =0$ subject to $V^{M}_A=H^{M}_{\xi=1}$ (top row) and  $V^{M}_B=H^{M}_{\xi=-1}$  (bottom row)
 sublattice selective moir\'e patterns. 
 {\bf a.}  Berry curvatures for valence and conduction flat bands for TLG/BN obtained for small interlayer potential differences of $\Delta = \pm0.0001$~eV giving rise to
 flat Chern bands with $C= \pm3$ and superlattice flat bands with $C=0$.
 {\bf b.} The Berry curvatures associated to bilayer BLG/BN with $\Delta = \pm 0.0001$~eV subject to $\xi=\pm1$ moir\'e potentials giving rise to superlattice flat bands with $C= \pm2$ and $C=0$.
 {\bf c.} The Berry curvatures associated to monolayer MLG/BN with $\Delta = \pm 0.01$~eV subject to sublattice selective $\xi=\pm1$ moir\'e potentials giving rise to superlattice flat bands with $C= \pm1$ and $C=0$.
}
\end{figure*}

\begin{figure*}
\begin{center}
\includegraphics[width=15cm]{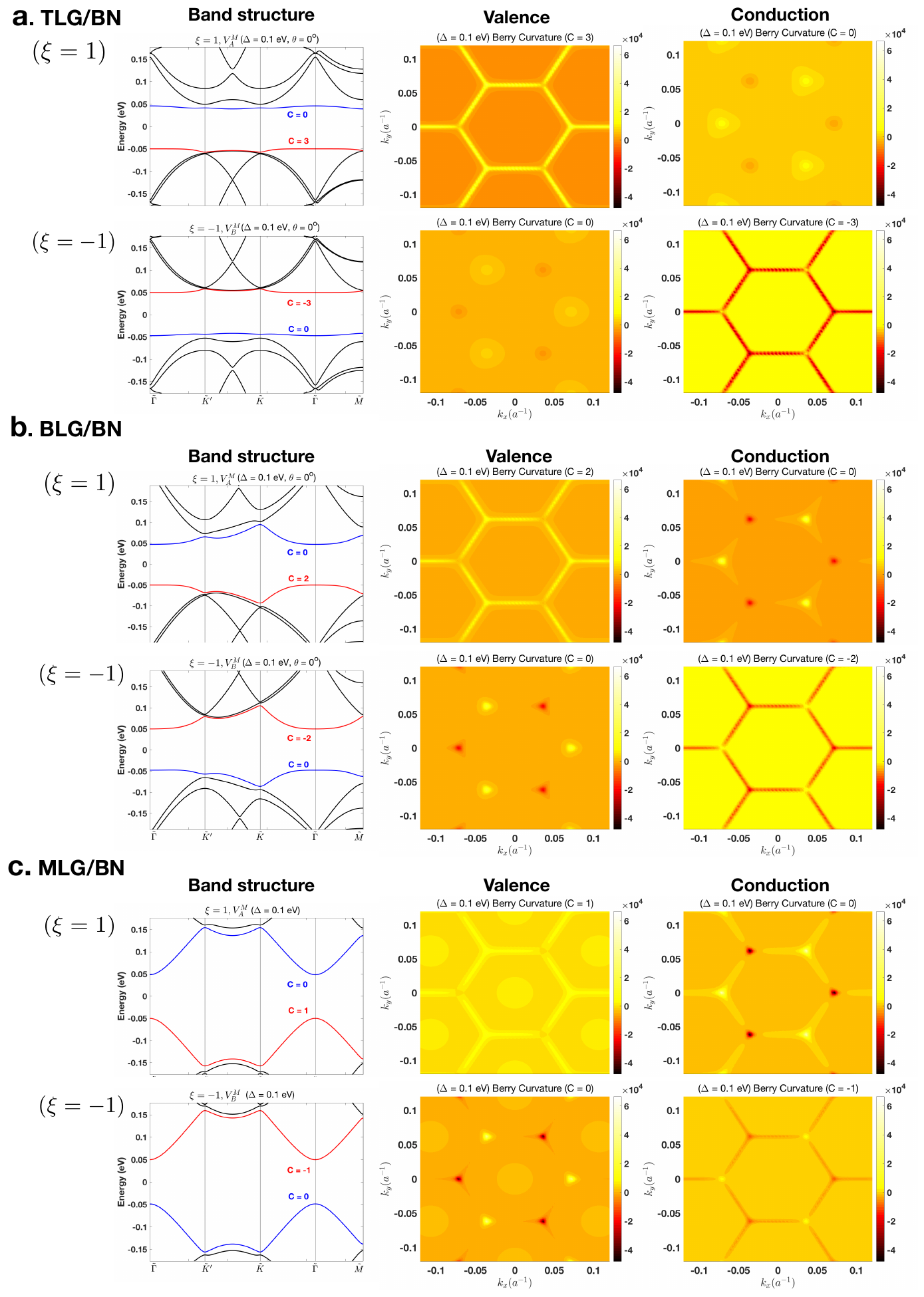}
\end{center}
 \caption{
 (Color online) The band structure of ABC-TLG ({\bf a. TLG/BN}), bilayer ({\bf b. BLG/BN}) and monolayer ({\bf c. MLG/BN}) 
 calculated using the minimal model without remote hopping terms (Eq.~1 in main text with $H^R_N=0$) for zero twist angle,  
 subject to $V^{M}_A=H^{M}_{\xi=1}$ and  $V^{M}_B=H^{M}_{\xi=-1}$ moir\'e patterns and interlayer 
 potential differences of $\Delta = 0.1$~eV  that give rise to flat Chern bands with $C= \pm 3$, for TLG/BN, $C= \pm 2$, for BLG/BN, and $C= \pm 1$ for MLG/BN. The Chern bands represented in red  and  the trivial $C=0$ bands are represented in blue.  We also plot the Berry curvatures associated to the valence  and conduction bands where we can observe how the Chern weights are pushed towards the mBZ boundaries. 
}
\end{figure*}

\begin{figure*}
\begin{center}
\includegraphics[width=15cm]{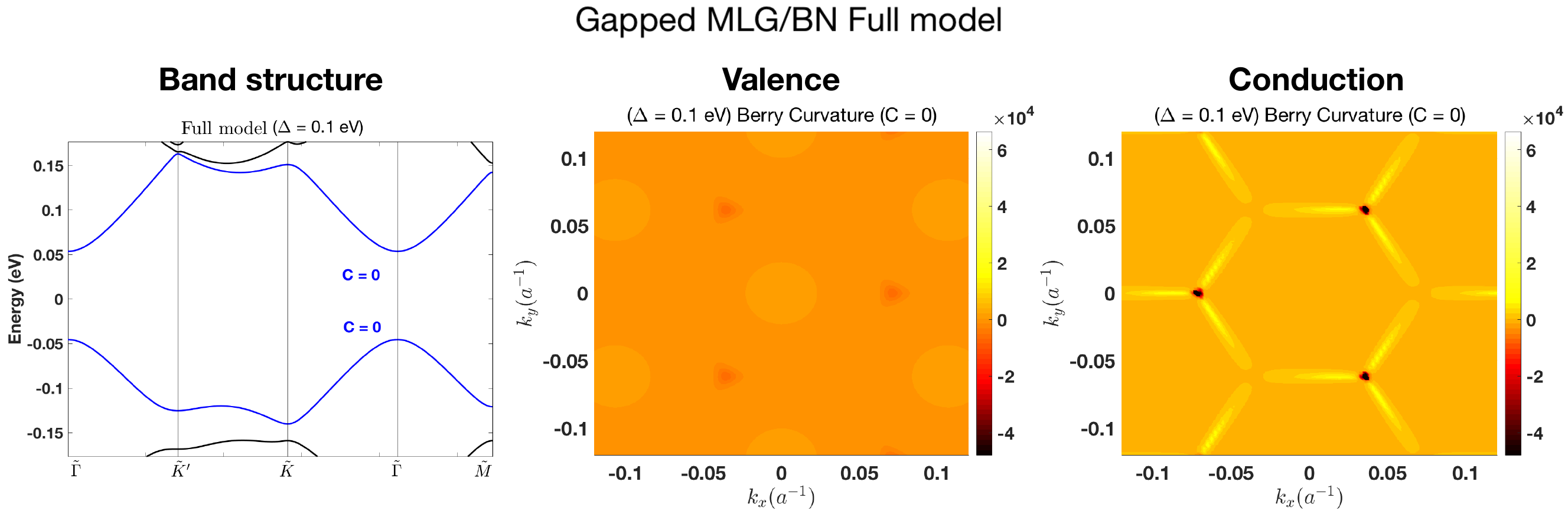}
\end{center}
 \caption{
 (Color online)  The band structure of the MLG/BN model with full moir\'e pattern potential including diagonal and off-diagonal terms 
 (PRB 89, 205414 (2014)) plus a primary gap $\Delta = 0.1$~eV  that leads to isolated valence and conduction bands with zero Chern number. 
The Berry curvatures for the valence and conduction bands show peaks near the mBZ boundaries and zone corners that cancel each other. 
}
\end{figure*}

\end{document}